\documentclass[cpc,twocolumn]{revtex4}%
\usepackage{amsfonts}
\usepackage{amsmath}
\usepackage{amssymb}
\usepackage{graphicx}
\begin{document}
\title{Harvesting entanglement from the cylindrical gravitational wave spacetime}
\author{Feifan He$^{1}$}
\author{Yongjie Pan$^{1}$}
\author{Baocheng Zhang$^{1}$}
\email{zhangbaocheng@cug.edu.cn}
\affiliation{$^{1}$School of Mathematics and Physics, China University of Geosciences,
Wuhan 430074, China}
\begin{abstract}
We investigate the entanglement harvesting protocol within the context of
cylindrical gravitational waves given first by Einstein and Rosen, focusing on
the interactions between non-relativistic quantum systems and linearized
quantum gravity. We study how two spatially separated detectors can extract
entanglement from the specific spacetime in the presence of gravitational
waves, which provides a precise quantification of the entanglement that can be
harvested using these detectors. In particular, we obtain the relation between
harvested entanglement and the distance to wave sources that emits
gravitational waves and analyze the detectability using quantum Fisher
information. The enhanced detectability demonstrates the advantages of
cylindrical symmetric gravitational waves.

\end{abstract}
\maketitle


\section{Introduction}

Entanglement harvesting~\cite{av91,br03}, refers to the process by which
detectors independently coupled to a quantum field can become entangled by
extracting entanglement from the field. This mechanism operates within a
multipartite quantum system framework, comprising the combined Hilbert spaces
of the detectors and the field. Typically modeled using a scalar field, this
setup facilitates the transfer of virtual particles between detectors, thereby
inducing entanglement among them. The possibility of entanglement harvesting
from spacelike separated regions is unique to quantum fields, as classical
fields do not possess entanglement that can be extracted. This distinction has
been utilized to determine the quantum or classical nature of a field.
Notably, it has been proposed that employing an entanglement harvesting
protocol with the gravitational field could serve as a direct witness to
quantum gravity~\cite{fpt20,pp21}. This approach underscores the pivotal role
of quantum field properties in facilitating such quantum phenomena.

Initially explored in flat spacetime scenarios~\cite{av91,br03}, this
phenomenon of entanglement harvesting has been extensively investigated under
various conditions, including cosmological backgrounds~\cite{sm09,mm12},
noninertial frames~\cite{smm15,lzy22}, black hole
environments~\cite{tm20,gtm21}, and in the presence of gravitational waves
(GWs)~\cite{xas20,gkt21}. Further studies have considered entanglement
harvesting when detectors interact with distinct field
operators~\cite{smm17,plm22,prm23} and when placed in superpositions of
different temporal orders or trajectories~\cite{hbm20,fmz21}.

The entanglement harvested by two Unruh-DeWitt (UDW) detectors
\cite{wgu76,bsd79} is highly sensitive to the frequency of the gravitational
wave. This can reveal the \textquotedblleft information
content\textquotedblright\ about gravitational-wave memory effect and
supertranslations \cite{xas20,gkt21}. Other investigations into entanglement
harvesting from the vacuum including gravitational waves involved the quantum
degrees of freedom of gravity by coupling GWs to a scalar quantum
field~\cite{ck11,cc22,tp23}. In these studies, the GWs were considered as
planar waves, which could manifest that the vacuum in the presence of GWs are
quantum, but cannot reveal any other information about the wave sources. In
this paper, we will investigate entanglement harvesting in the context of
cylindrical GWs of Einstein and Rosen (called also Einstein-Rosen waves, or
ERWs) \cite{er37,ww57,bgp19,hz22} and discuss the information about the
distance from the wave sources obtained through the analyses on the
entanglement change between two UDW detectors.

The ERW, an exact solution to general relativity characterized by two
commuting Killing vectors, aptly describes a cylindrical GW. Historically, the
ERW was pivotal in early explorations to quantify the energy transported by
GWs~\cite{kst65,sc86,mt17,bsw20}, a challenging task due to the local
non-descriptiveness of GW energy caused by the equivalence
principle~\cite{rai68,mab97}, which showed that the observation of ERWs is
feasible. Moreover, the quantum aspects of the ERWs have been rigorously
formulated~\cite{kk71}, and its quantization in conjunction with a massless
scalar field has been successfully achieved~\cite{bgv05}. This facilitates the
study about entanglement harvesting from the ERW spacetime. In particular, the
ERWs carry the information about the distance from the wave sources, which
might be transferred to the harvested entanglement between the detectors, as
will be investigated in this paper.

This paper is organized as follows. In Section II, the protocol of
entanglement harvesting is revisited using two spacelike separated detectors
that are accelerating in the flat spacetime. In Section III, the quantized
formalism of the weak cylindrical GWs, and how the detectors are coupled to
the GWs are described. Meanwhile, entanglement harvesting from the spacetime
in the presence of linear cylindrical GWs is studied and the information about
the distance from the wave sources is revealed in this section. The
conclusions is given in Section IV.

\section{Entanglement Harvesting Protocol}

\subsection{The Untuh-DeWitt detector}

We consider two identical UDW detectors, labeled $A$ and $B$. Each detector is
a two-level system with a ground state $\left\vert g\right\rangle _{D}$ and an
excited state $\left\vert e\right\rangle _{D}$, where $D\in\{{A,B\}}$. The
energy difference between these states is denoted by $\Omega_{D}$. These
detectors interact locally with a massless quantum scalar field, represented
as $\hat{\phi}(x,t)$. The path followed by each detector through spacetime is
specified by $x_{D}(\tau_{D})$, where $\tau_{D}$ is the proper time
experienced by the detector $D$. The interaction between each detector and the
scalar field is governed by a Hamiltonian specific to each detector,
\begin{equation}
H_{D}(\tau)=\lambda_{D}\chi_{D}(\tau)\left(  f_{D}^{\ast}(x)e^{i\Omega_{D}%
\tau}\sigma_{D}^{+}+f_{D}(x)e^{-i\Omega_{D}\tau}\sigma_{D}^{-}\right)
\phi(x), \label{iH}%
\end{equation}
where $\lambda_{D}\ll1$ represents a small coupling strength of each detector
to the scalar field. The switching function, $\chi_{D}(\tau)$, regulates the
timing of the interaction, turning the coupling to the field on and off.
$f_{D}(x)$ is the smearing function which controls the spatial region of the
interaction. The ladder operators, which facilitate transitions between the
detector states, are defined as $\sigma_{D}^{+}=\left\vert e\right\rangle
_{D}\left\langle g\right\vert _{D}$ and $\sigma_{D}^{-}=\left\vert
g\right\rangle _{D}\left\langle e\right\vert _{D}$. These operators are
crucial in our quantum mechanical model, which effectively describes
light-matter interactions without involving angular momentum exchange.

The time evolution of the detector-field system is governed by the unitary
operator $\hat{U}$, defined as
\begin{equation}
\hat{U}=\mathcal{T}\exp\left[  -%
{\displaystyle\int}
dt\left(  \frac{d\tau_{A}}{dt}\hat{H}_{A}[\tau_{A}(t)]+\frac{d\tau_{B}}%
{dt}\hat{H}_{B}[\tau_{B}(t)]\right)  \right]  \label{tv}%
\end{equation}
where $\mathcal{T}$ is the time-ordering operator that arranges the operators
from earliest to latest times as we move from right to left in the exponential.

Initially, both detectors $A$ and $B$ start in their ground states, and the
field is in its vacuum state. The combined initial state of the system is
\begin{equation}
\left\vert \varphi_{0}\right\rangle =\left\vert g\right\rangle _{A}%
\otimes\left\vert g\right\rangle _{B}\otimes\left\vert 0\right\rangle _{\phi}.
\end{equation}

After the interaction, the state of the detectors is described by a density
matrix $\hat{\rho}_{AB}=Tr_{\phi}\left[  \hat{U}\left\vert \varphi
_{0}\right\rangle \left\langle \varphi_{0}\right\vert \hat{U}^{\dagger
}\right]  $, obtained by tracing out the field states from the total system
state. This results in
\begin{equation}
\hat{\rho}_{AB}=%
\begin{pmatrix}
1-\mathcal{L}_{AA}-\mathcal{L}_{BB} & 0 & 0 & \mathcal{M}^{\ast}\\
0 & \mathcal{L}_{BB} & \mathcal{L}_{BA} & 0\\
0 & \mathcal{L}_{AB} & \mathcal{L}_{AA} & 0\\
\mathcal{M} & 0 & 0 & 0
\end{pmatrix}
+\mathcal{O}(\lambda^{2}) \label{SC}%
\end{equation}
to lowest order in the coupling strength. The density matrix (\ref{SC}) is
expressed in the basis \{$\left\vert g_{A}g_{B}\right\rangle ,\left\vert
g_{A}e_{B}\right\rangle ,\left\vert e_{A}g_{B}\right\rangle ,\left\vert
e_{A}e_{B}\right\rangle $\}. Here $\mathcal{L}_{AA}$ and $\mathcal{L}_{BB}$
represents the probability of detector being excited, and $\mathcal{L}_{AB}$,
$\mathcal{L}_{BA}$, and $\mathcal{M}$ measures the coherence between the
detectors due to their interaction with the field, reflecting nonlocal effects
between the two detectors at different times. Their expressions are given as
\begin{align}
\mathcal{L}_{IJ}  &  =\lambda_{I}\lambda_{J}%
{\displaystyle\int}
d\tau_{I}d\tau_{J}\chi_{I}\left(  \tau_{I}\right)  \chi_{J}\left(  \tau
_{J}\right)  f_{I}(x_{I})f_{J}^{\ast}(x_{J})\nonumber\\
&  \times e^{-\left(  \Omega_{I}\tau_{J}-\Omega_{I}\tau_{J}\right)  }W\left(
x_{I}(t),x_{J}(t^{\prime})\right)
\end{align}
and
\begin{align}
\mathcal{M}  &  =\lambda_{A}\lambda_{B}%
{\displaystyle\int}
d\tau_{A}d\tau_{B}\chi_{A}\left(  \tau_{A}\right)  \chi_{B}\left(  \tau
_{B}\right)  f_{A}(x_{A})f_{B}^{\ast}(x_{B})\nonumber\\
&  \times e^{-\left(  \Omega_{A}\tau_{A}+\Omega_{B}\tau_{B}\right)  }%
\theta(t^{\prime}-t)\nonumber\\
&  \times(W(x_{A}(t),x_{B}(t^{\prime}))+W(x_{B}(t),x_{A}(t^{\prime}))
\end{align}
where $\theta(t-t^{\prime})$ is the Heaviside function. The Wightman function
$W(x,x^{\prime})=\langle0|\hat{\phi}\left(  x(t),t\right)  ,\hat{\phi}\left(
x^{\prime}(t^{\prime}),t^{\prime}\right)  |0\rangle$ is a fundamental field
correlator that quantifies the vacuum fluctuations of the field between two
spacetime points $x(t)$ and $x^{\prime}(t^{\prime})$, crucial for
understanding the field's influence on the detectors.

When we consider only one of the detectors, either $A$ or $B$, by tracing out
the other one from the combined density matrix (Eq.\ \ref{SC}), we arrive at a
simplified description for the state of the remaining detector,
\begin{equation}
\hat{\rho}_{D}=%
\begin{pmatrix}
1-\mathcal{L}_{C} & 0\\
0 & \mathcal{L}_{C}%
\end{pmatrix}
\quad.
\end{equation}
In this matrix, $\mathcal{L}_{C}$ represents the probability that the detector
$A$ or $B$ transitions from its ground state to its excited state due to its
interaction with the field.

\subsection{Negativity}

To explore the entanglement between two identical UDW detectors after local
interactions with a quantum field, we use a measure called negativity
\cite{vw02,mbp05}. Negativity is a reliable quantifier for entanglement
between two qubits, suitable for situations such as entanglement harvesting as
discussed in previous studies. The negativity of a system, $\mathcal{N}$, is
determined by summing the negative eigenvalues from the partial transpose of
the density matrix $\hat{\rho}_{D}$,
\begin{equation}
\mathcal{N}=\text{max}\left(  0,\sqrt{|\mathcal{M}|^{2}-\frac{(\mathcal{L}%
_{AA}-\mathcal{L}_{BB})^{2}}{4}}-\frac{\mathcal{L}_{AA}+\mathcal{L}_{BB}}%
{2}\right)  . \label{Ne}%
\end{equation}
In cases where the detectors have equal excitation probabilities,
$\mathcal{L}_{AA}=\mathcal{L}_{BB}=\mathcal{L}$, the formula simplifies to
\begin{equation}
\mathcal{N}=\text{max}(0,\,|\mathcal{M}|-\mathcal{L}).
\end{equation}
Assuming both detectors are identical with equal interaction strengths,
frequencies, and simultaneous interactions in their respective frames, we can
use the following expressions to compute the necessary probabilities and
correlation terms from the Fourier transforms of the switching functions,
\begin{align}
\mathcal{L}_{IJ}\!  &  =\frac{\lambda^{2}}{(2\pi)^{3}}\int\frac{d^{3}%
\mathbf{k}}{2|\mathbf{k}|}\tilde{\chi}^{\ast}(\Omega+|\mathbf{k}|)\tilde{\chi
}(\Omega+|\mathbf{k}|)\tilde{f}_{I}(\mathbf{k})\tilde{f}_{J}^{\ast}%
(\mathbf{k}),\label{Scl}\\
\mathcal{M}\!  &  =\!-\frac{\lambda^{2}}{(2\pi)^{3}}\!\!\!\int\!\!\frac
{d^{3}\mathbf{k}}{2|\mathbf{k}|}Q(|\mathbf{k}|,\Omega)(\Tilde{f}%
_{A}(\!-\mathbf{k})\Tilde{f}_{B}(\mathbf{k})\!+\!\Tilde{f}_{B}(\!-\mathbf{k}%
)\Tilde{f}_{A}(\mathbf{k})), \label{Scm}%
\end{align}
where
\begin{equation}
Q(|\mathbf{k}|,\Omega)=\int dtdt^{\prime}\chi(t)\chi(t^{\prime})e^{i(\Omega
+|\mathbf{k}|)t^{\prime}}e^{i(\Omega-|\mathbf{k}|)t}\theta(t-t^{\prime}),
\label{Q}%
\end{equation}
and we have defined the Fourier transform of $f(\mathbf{x})=\psi
_{e}(\mathbf{x})\psi_{g}^{\ast}(\mathbf{x})$ and $\chi(t)$ as
\begin{align}
\tilde{f}(\mathbf{k})  &  =\int d^{3}\mathbf{x}\,f(\mathbf{x})e^{i\mathbf{k}%
\cdot\mathbf{x}},\label{fF}\\
\tilde{\chi}(\omega)  &  =\int dt\,\chi(t)e^{i\omega t}. \label{fC}%
\end{align}
It is noted that the results are obtained using the assumption that
$\lambda_{A}=\lambda_{B}=\lambda$, $\Omega_{A}=\Omega_{B}=\Omega$, $\chi
_{A}(t)=\chi_{B}(t)=\chi(t)$, and the smearings are identical modulo a spatial
translation. In the later calculations in this paper, this assumption will be maintained.

The concept of entanglement harvesting using two UDW detectors linearly
coupled to a scalar quantum field has been investigated in the literature
\cite{lzy22,tm20,plm22,xas20,klm17,hhz18,shk21,fmz21,gpm22}. But in the curved
spacetime, what information about the curved spacetime can be obtained from
the harvested entanglement was hardly investigated. This is the task of this
paper, and in what follows we will investigate how to extract the information
about the distance from the wave sources by the harvested entanglement from
the vacuum of the cylindrical GW spacetime.

\section{Einstein-Rosen Waves}

In this section we study the situation in which two UDW detectors are coupled
to the cylindrical GWs. We are going to implement the entanglement harvesting
protocol and analyze the information about the distance from the wave sources
using Fisher information{. }

\subsection{Quantized Cylindrical Gravitational Waves}

Since the observable effect of cylindrical GWs is considered at the place with
a large distance from the source, the linearized metric is adequate for our
purpose with the form as%
\begin{equation}
ds^{2}=\left(  1-\psi\right)  ds_{3}^{2}+\left(  1+\psi\right)  dZ^{2},
\label{metric}%
\end{equation}
where $ds_{3}^{2}=-\left(  1+\gamma\right)  dT_{C}^{2}+\left(  1+\gamma
\right)  dR^{2}+R^{2}d\theta^{2}$, and $\psi$ and $\gamma$ are functions of
only $R$ and $T_{C}$. It derives from the spacetime metric of ERWs
\cite{er37,kk71,rt96,cor16}, $ds^{2}=e^{\gamma-\psi}(-dT_{C}^{2}%
+dR^{2})+e^{-\psi}R^{2}d\theta^{2}+e^{\psi}dZ^{2}$, in which $\psi$ encodes
the physical degrees of freedom and satisfies the usual wave equation for an
axially symmetric massless scalar field in three-dimensions,
\begin{equation}
\partial_{T_{C}}^{2}\psi-\partial_{R}^{2}\psi-\frac{1}{R}\partial_{R}\psi=0\,.
\end{equation}
The metric function $\gamma$ can be expressed as \cite{ap96}, $\gamma
(R)=\frac{1}{2}\int_{0}^{R}d\bar{R}\,\bar{R}\,\left[  (\partial_{T_{C}}%
\psi)^{2}+(\partial_{\bar{R}}\psi)^{2}\right]  ,$and $\gamma_{\infty}=\frac
{1}{2}\int_{0}^{\infty}dR\,R\,\left[  (\partial_{T_{C}}\psi)^{2}+(\partial
_{R}\psi)^{2}\right]  $, where $\gamma(R)$ and $\gamma_{\infty}$ are the
energy of the scalar field in a ball of radius $R$ and in the whole
two-dimensional flat space, respectively.

When the regularity at the origin $R=0$ is imposed \cite{ap96}, the solutions
for the field $\psi$ can be expanded in the form
\begin{equation}
\psi(R,T_{C})=\int_{0}^{\infty}\!\!\frac{d\mathbf{k}}{\sqrt{2}}\,J_{0}%
(R\mathbf{k})\left[  A(\mathbf{k})e^{-ikT_{C}}+A^{\dagger}(\mathbf{k}%
)e^{ikT_{C}}\right]  , \label{qerw}%
\end{equation}
where $A(\mathbf{k})$ and $A^{\dagger}(\mathbf{k})$ are fixed by the initial
conditions and are complex conjugate to each other, because $\psi$ and $J_{0}$
(the zeroth-order Bessel function of the first kind) are real. In principle,
the quantization of the field $\psi$ can be carried out in a standard way. We
can introduce a Fock space in which $\hat{\psi}(R,0)$, the quantum counterpart
of $\psi(R,0)$, is an operator-valued distribution \cite{bs75}. Its action is
determined by the usual annihilation and creation operators, $\hat
{A}(\mathbf{k})$ and $\hat{A}^{\dagger}(\mathbf{k})$, whose only non-vanishing
commutators are $\left[  \hat{A}(\mathbf{k}_{1}),\hat{A}^{\dagger}%
(\mathbf{k}_{2})\right]  =\delta(\mathbf{k}_{1},\mathbf{k}_{2}).$

The Hamiltonian of this linearized gravity can be written as
\cite{kk71,rt96,bmv03},%
\begin{equation}
H_{0}=\int_{0}^{R}dR\left(  \frac{p_{\hat{\psi}}^{2}}{2R}+\frac{R}{2}\left(
\frac{\partial\hat{\psi}}{\partial R}\right)  \right)  ,
\end{equation}
where the gauge fixing conditions $p_{\gamma}=0$ and $R=r$. $p_{\hat{\psi}}$
and $p_{\gamma}$ are the canonical momenta conjugated to the metric fields
$\hat{\psi}$ and $\gamma$, respectively. $R=r$ indicates that $R$ can be used
to measure the distance from the source to the detector. It is not hard to
confirm that $H_{0}=\gamma_{\infty}=\int_{0}^{\infty}d\mathbf{k}%
\,\mathbf{k}\hat{A}^{\dagger}(\mathbf{k})\hat{A}(\mathbf{k})$ when the
expression of $p_{\hat{\psi}}$ is used.

To get a unit asymptotic timelike Killing vector field in the actual
four-dimensional spacetime \cite{av94,mv95}, one must transfer the time
coordinate by $T_{C}=e^{-\gamma_{\infty}/2}t$. In this asymptotic region
$R\rightarrow\infty$, $\partial_{t}$ is a unit timelike vector. $t$ is the
physical time, and the corresponding physical Hamiltonian is given as
\cite{rt96,av94,mv95}, $H=E(H_{0})=2(1-e^{-H_{0}/2})$. Thus, the annihilation
operator with respect to the physical time can be linked to $\hat
{A}(\mathbf{k})$ by $\hat{A}_{E}(\mathbf{k},t)=\hat{A}(\mathbf{k}%
)\exp[-itE(\mathbf{k})e^{-H_{0}/2}]$. In the first approximation, the physical
field $\hat{\psi}=$ $\int_{0}^{\infty}\!\!\frac{d\mathbf{k}}{\sqrt{2}}%
\,J_{0}(R\mathbf{k})\left[  \hat{A}_{E}(\mathbf{k},t)+\hat{A}_{E}^{\dagger
}(\mathbf{k},t)\right]  $ has the similar time-evolved form to Eq.
(\ref{qerw}). When the UDW detectors are coupled to the linearized cylindrical
GWs, the physical Hamiltonian $H$ should be considered, but in the first-order
perturbation, $H\simeq H_{0}$ and $t\simeq T_{C}$. Thus, the results in the
coordinates ($T_{C},R,\theta,Z$) can be used in the actual interaction between
the detectors and the linearized cylindrical GWs, as seen in the next section.

\subsection{Entanglement Harvesting}

We start with the interaction Hamiltonian between gravitational waves and two
free falling detectors \cite{prm23}%
\begin{equation}
\hat{H}_{I}(t)=\lambda\mathcal{R}_{0i0j}(t,\hat{\mathbf{x}})\hat{x}^{i}\hat
{x}^{j},
\end{equation}
where $\lambda=\sqrt{\frac{\pi}{2}}\frac{m}{m_{p}}$, with $m_{p}$ being the
Planck mass. Here, $\lambda$ serves as a dimensionless coupling constant,
essentially scaling with the detector's rest mass measured in Planck units.
This formulation allows us to quantitatively assess the gravitational effects
on quantum mechanical scales. There are some other ways (see the discussions
in Ref. \cite{prm23}) to describe the interaction between the detectors and an
external weak gravitational field, while we choose the same way as in Ref.
\cite{prm23} which could be obtained by considering a wave function in curved
spacetimes, since for our study the quantum states for the detectors and the
quantum description for ERWs are explicit as presented in the following calculations.

Assuming that the energy levels of our detector's free Hamiltonian are
discrete, we can expand the interaction Hamiltonian in terms of the system's
wavefunctions as
\begin{align}
\hat{H}_{I}(t)\!  &  =\lambda\!\chi(t)\int\!\!d^{3}\mathbf{x}\mathcal{R}%
_{0i0j}(t,\mathbf{x}){x}^{i}{x}^{j}\left\vert \mathbf{x}\right\rangle
\left\langle \mathbf{x}\right\vert \!\!_{t}\nonumber\\
&  =\!\lambda\chi(t)\sum_{nm}\!\int\!d^{3}\mathbf{x}\mathcal{R}_{0i0j}%
(t,\mathbf{x}){x}^{i}{x}^{j}f_{nm}^{\ast}(\mathbf{x})e^{\mathrm{i}\Omega
_{nm}t}\left\vert n\right\rangle \left\langle m\right\vert , \label{HN}%
\end{align}
where $f_{nm}(\mathbf{x})=\psi_{n}(\mathbf{x})\psi_{m}^{\ast}(\mathbf{x})$ is
the smearing function, $\left\vert \mathbf{x}\right\rangle \left\langle
\mathbf{x}\right\vert _{t}$ denotes the position operator in the interaction
picture, and the switching function is added here in order to ensure the
finite interaction time. The functions $\psi_{n}(\mathbf{x})=\left\langle
\mathbf{x}|n\right\rangle $ represent the wavefunctions corresponding to the
energy eigenvalues $E_{n}$, and $\Omega_{nm}=E_{n}-E_{m}$ represents the
energy difference between states. In our calculation, the detectors are
regarded as two-level atoms, such as the ground state and an excited state,
and the respective wavefunctions are $\psi_{g}(\mathbf{x})=\left\langle
\mathbf{x}|g\right\rangle $ for the ground state and $\psi_{e}(\mathbf{x}%
)=\left\langle \mathbf{x}|e\right\rangle $ for the excited state. This
simplified model allows us to focus on the key dynamical aspects of the
quantum system under the influence of an external gravitational field. Then,
the interaction Hamiltonian can be written as
\begin{align}
\hat{H}_{I}(t)=\lambda\chi(t)\int d^{3}\mathbf{x}\,(F^{ij\ast}(\mathbf{x}%
)e^{i\Omega t}  &  \hat{\sigma}^{+}+F^{ij}(\mathbf{x})e^{-i\Omega t}%
\hat{\sigma}^{-})\nonumber\label{eq:GravInteraction}\\
&  \times\mathcal{R}_{0i0j}(t,\mathbf{x}),
\end{align}
where the energy difference between the excited state $\left\vert
e\right\rangle $ and the ground state $\left\vert g\right\rangle $ is denoted
by $\Omega=\Omega_{eg}=E_{e}-E_{g}$. The function $F^{ij}(\mathbf{x})=\psi
_{e}(\mathbf{x})\psi_{g}^{\ast}(\mathbf{x})x^{i}x^{j}$ represents the smearing
tensors, which are crucial for modeling the interaction of the detector with
the linearized gravitational field. The ladder operators are defined as
$\hat{\sigma}^{+}=\left\vert e\right\rangle \left\langle g\right\vert $, and
$\hat{\sigma}^{-}=\left\vert g\right\rangle \left\langle e\right\vert $.
Additionally, the terms in the Hamiltonian that commute with the detector's
free Hamiltonian have been neglected, as they do not contribute to the
entanglement dynamics but only shift the energy levels. For the quantization,
it can be implemented by replacing the curvature tensor $\mathcal{R}%
_{0i0j}(t,\mathbf{x})$ with the operator-valued distribution $\hat
{\mathcal{R}}_{0i0j}(x)$ in the Hamiltonian. This model provides a framework
for understanding the interaction between a localized quantum system and a
weak quantum gravitational field.

To leading order in $\lambda$, the excitation probability of the detector
after the interaction can be expressed as
\begin{align}
\mathcal{L}^{G}=\lambda^{2}\int d^{4}xd^{4}x^{\prime}\chi(t)\chi(t^{\prime})
&  F^{ij}(\mathbf{x})F^{kl\ast}(\mathbf{x}^{\prime})e^{-i\Omega(t-t^{\prime}%
)}\nonumber\\
&  \times\langle\hat{\mathcal{R}}_{0i0j}(x)\hat{\mathcal{R}}_{0k0l}(x^{\prime
})\rangle_{0}.
\end{align}
This equation shows how the excitation probability can be transformed into a
single momentum integral based on the curvature two-point function. This
formulation is crucial for quantitatively describing how quantum systems
respond to gravitational fields, offering insights into their probabilistic
behavior under such influences. The curvature fluctuation which is given by
$R_{0R0R}(R,T)=-\frac{1}{2}\partial_{T}^{2}h_{RR}$ ($h_{RR}=\psi\left(
R,T\right)  $ as seen in Eq. (\ref{metric})) can be obtained as
\begin{align}
\hat{\mathcal{R}}_{0R0R}(R,T)  &  =\int\frac{d\mathbf{k}}{2\sqrt{2}%
}|\mathbf{k}|^{2}J_{0}(\mathbf{k}R)\nonumber\\
&  \times\left[  \hat{A}(\mathbf{k})e^{-ikT}+\hat{A}^{\dagger}(\mathbf{k}%
)e^{ikT}\right]  , \label{QC}%
\end{align}
where Eq. (\ref{qerw}) is used. Then, the curvature two-point function is
calculated as
\begin{align}
&  \left\langle \hat{\mathcal{R}}_{0R0R}\left(  R\right)  \hat{\mathcal{R}%
}_{0R0R}(R^{\prime})\right\rangle _{0}\nonumber\\
&  =%
{\displaystyle\int}
\frac{d\mathbf{k}}{8}\left\vert \mathbf{k}\right\vert ^{4}J_{0}\left(
\mathbf{k}R\right)  J_{0}\left(  \mathbf{k}R^{\prime}\right)  e^{-ik\left(
T-T^{\prime}\right)  }.
\end{align}

Thus,%

\begin{align}
\mathcal{L}^{G}  &  =\lambda^{2}\frac{4T^{2}\sigma^{8}}{15\pi}\int
d|\boldsymbol{k}|\mathrm{~}|\boldsymbol{k}|^{10}J_{0}(\boldsymbol{k}%
R)J_{0}(\boldsymbol{k}(R-L))\nonumber\\
&  \times e^{-|\boldsymbol{k}|^{2}\sigma^{2}}e^{-T^{2}(|\boldsymbol{k}%
|+\Omega)^{2}}.
\end{align}
where in the calculation the switching functions are taken as $\chi
_{A}(t)=\chi_{B}(t)=$ $\chi(t)=\frac{1}{\sqrt{2\pi}}\exp(-t^{2}/2T^{2})$, and
the smearing functions are takens as $f_{A}(x)=\frac{1}{(2\pi\sigma^{2}%
)^{3/2}}\exp(-x^{2}/2\sigma^{2})$, $f_{B}(x)=\frac{1}{(2\pi\sigma^{2})^{3/2}%
}\exp(-(x-L)^{2}/2\sigma^{2})$ where $T$ represents a duration timescale,
$\sigma$ determines the spatial width of the smearing function, and $L$ is the
separation between the detectors.

Similarly, given the choices of gaps and spacetime smearing functions, the
resulting expression for the non-local term $\mathcal{M}^{G}$ can be obtained
as
\begin{align}
\mathcal{M}^{G}  &  =-\lambda^{2}\frac{4T^{2}\sigma^{8}}{L^{5}\pi}%
\!\!\!\int\!\!d|\mathbf{k}|\,\,|\mathbf{k}|^{5}J_{0}(\mathbf{k}R)J_{0}%
(\mathbf{k}(R-L))e^{-|\mathbf{k}|^{2}\sigma^{2}}\nonumber\\
&  \times e^{-T^{2}(|\mathbf{k}|^{2}+\Omega^{2})}(1-\text{erf}[i|\mathbf{k}%
|T)(3|\mathbf{k}|L\cos(|\mathbf{k}|L)\nonumber\\
&  +(|\mathbf{k}|^{2}L^{2}-3)\sin(|\mathbf{k}|L)].
\end{align}
In Figs. 1, 2 and 3, we present the numerical results for the negativity.

\begin{figure}[h]
\centering
\includegraphics[width=8.6cm]{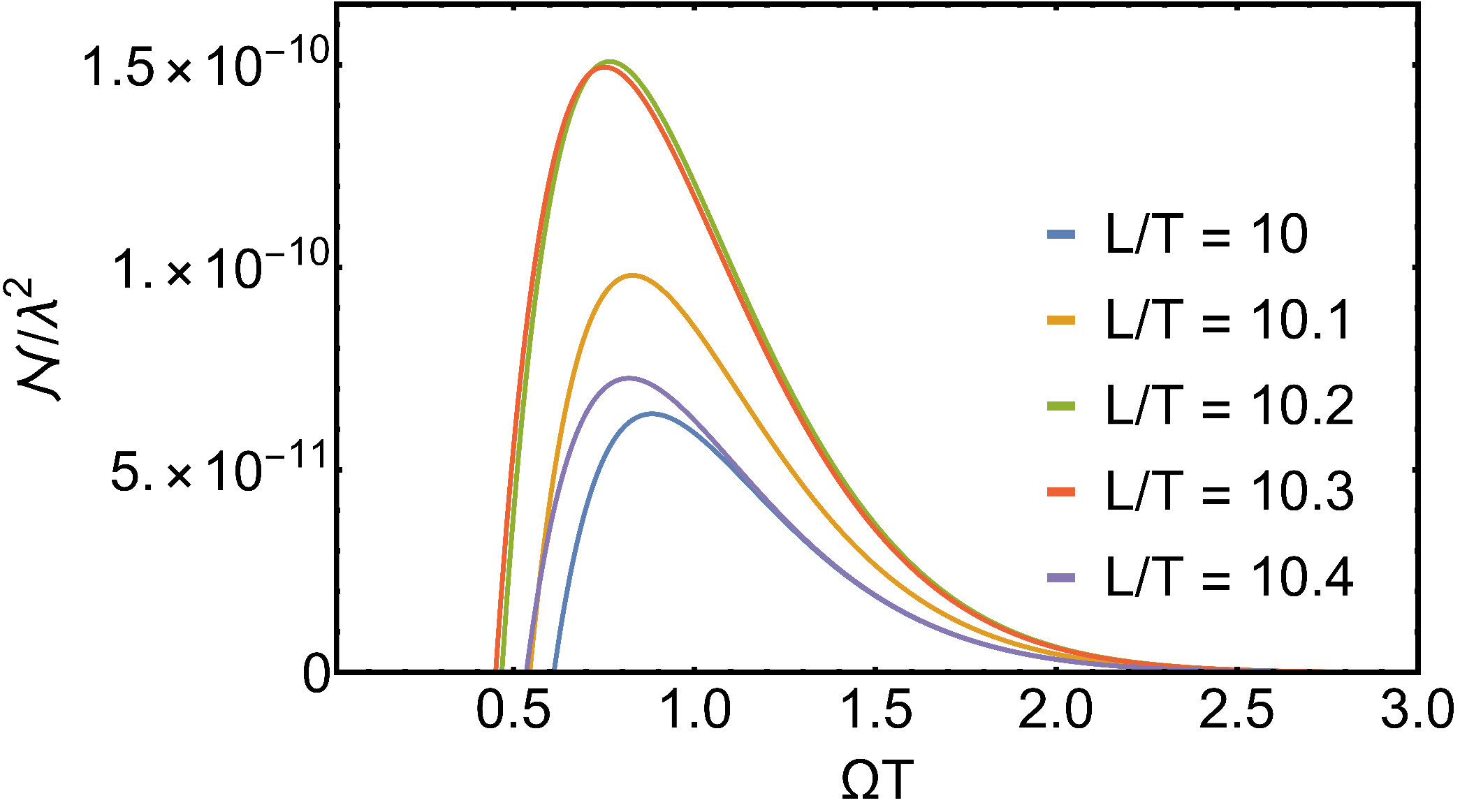} \caption{Negativity as a function of
the detectors' gap $\Omega$ for multiple values of the detectors separation
distance $L$. We fixed the detectors size to be $\sigma=0.3T$ for each of the
plots. Other parameter is taken as $R/T=1000$. }%
\label{Fig1}%
\end{figure}

In Fig. 1, we plot the negativity of the two-detector system as a function of
the detectors' energy gap, for different values of the separation between
them. We see that there is a minimum threshold on the required energy gap
before any entanglement is acquired between the detectors. Once the threshold
energy gap is met, there is a rapid increase in the negativity, until it
peaks. This is a similar behavior to entanglement harvesting from a real
scalar field, where the detectors gap can be tuned to maximize the harvested
entanglement (see, \cite{gpm22}).

\begin{figure}[h]
\includegraphics[width=8.6cm]{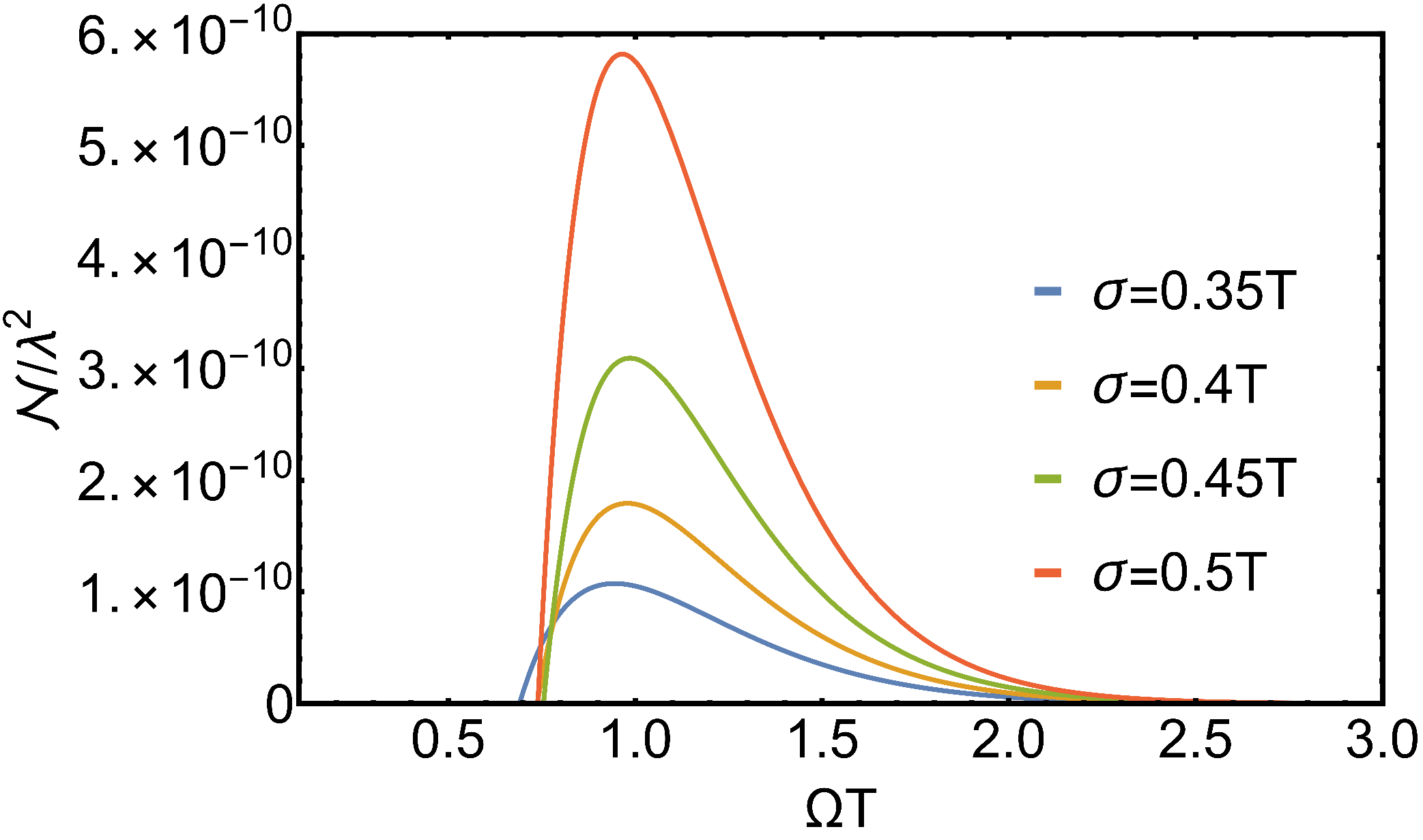} \caption{Negativity as a function of
the detectors' gap $\Omega$ for multiple values of the detectors' size
$\sigma$. We fixed the separation between the detectors to be $L = 10 T$ for
each of the plots. Other parameters is taken as $R/T=1000$. }%
\label{Fig2}%
\end{figure}

\begin{figure}[h]
\includegraphics[width=8.6cm]{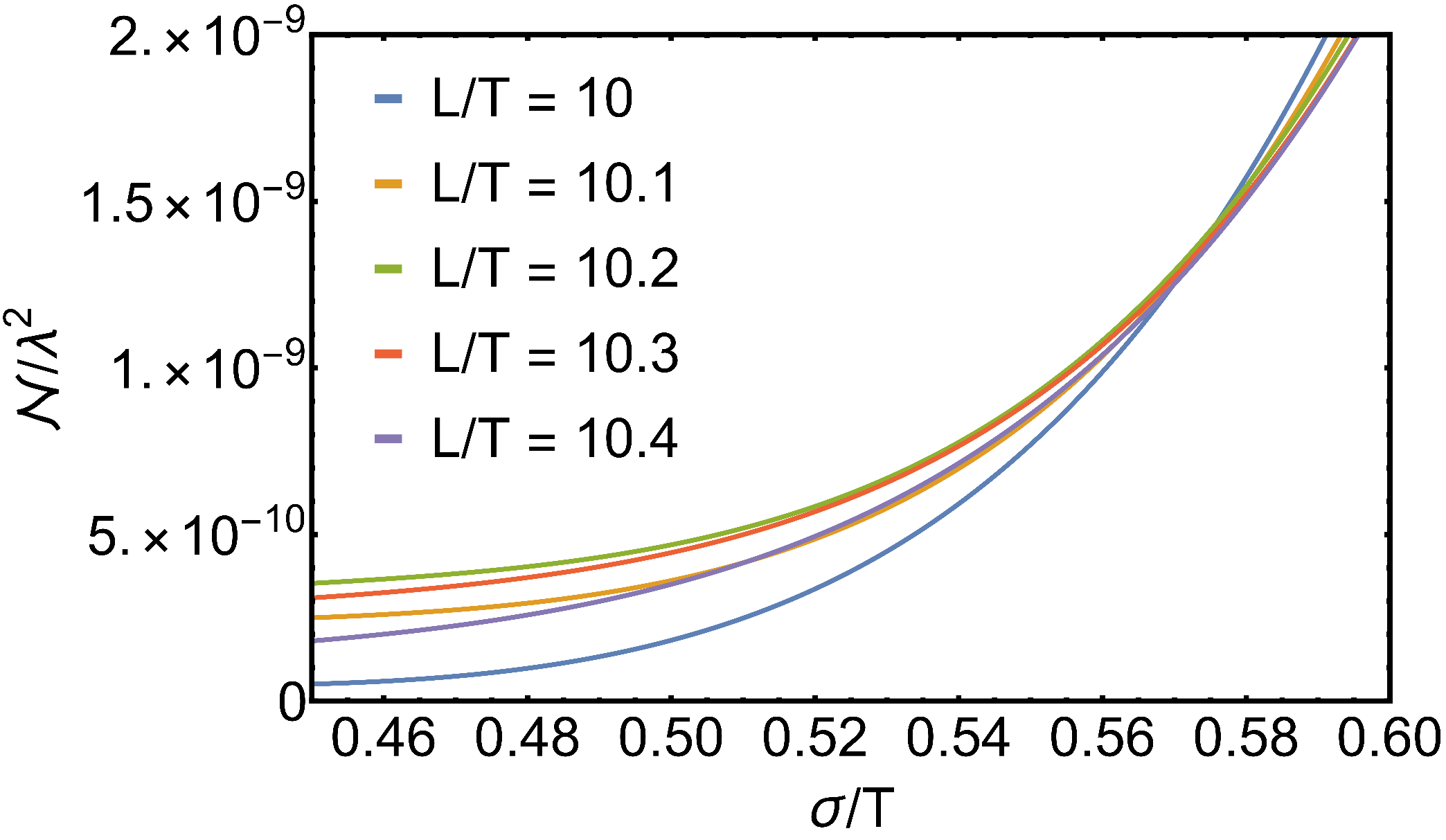} \caption{{Negativity as a function of
the detectors size $\sigma$ for multiple values of the detectors separation
$L$. We fixed the energy gap of the detectors as $\Omega T = 0.77$ for each of
the plots. Other parameter is taken as $R/T=1000$.}}%
\label{Fig3}%
\end{figure}

In Fig 2, we plot the entanglement acquired by the detectors as a function of
their energy gaps for varying detector sizes. We conclude that as the
detectors increase in size, the harvested entanglement increases. This can be
traced back to the fact that the interaction of the detectors with the
gravitational field is proportional to their sizes squared.

In Fig.~3, we plot the negativity of the detectors state as a function of
$\sigma$ for a fixed $\Omega T$ and varying values of $L$. We clearly see a
monotonic increase in the negativity with $\sigma$. We also see the result
that the negativity decreases only after the ratio $\sigma/T$ exceeds about
$0.57$ as the separation between the detectors increases.

\begin{figure}[h]
\includegraphics[width=8.6cm]{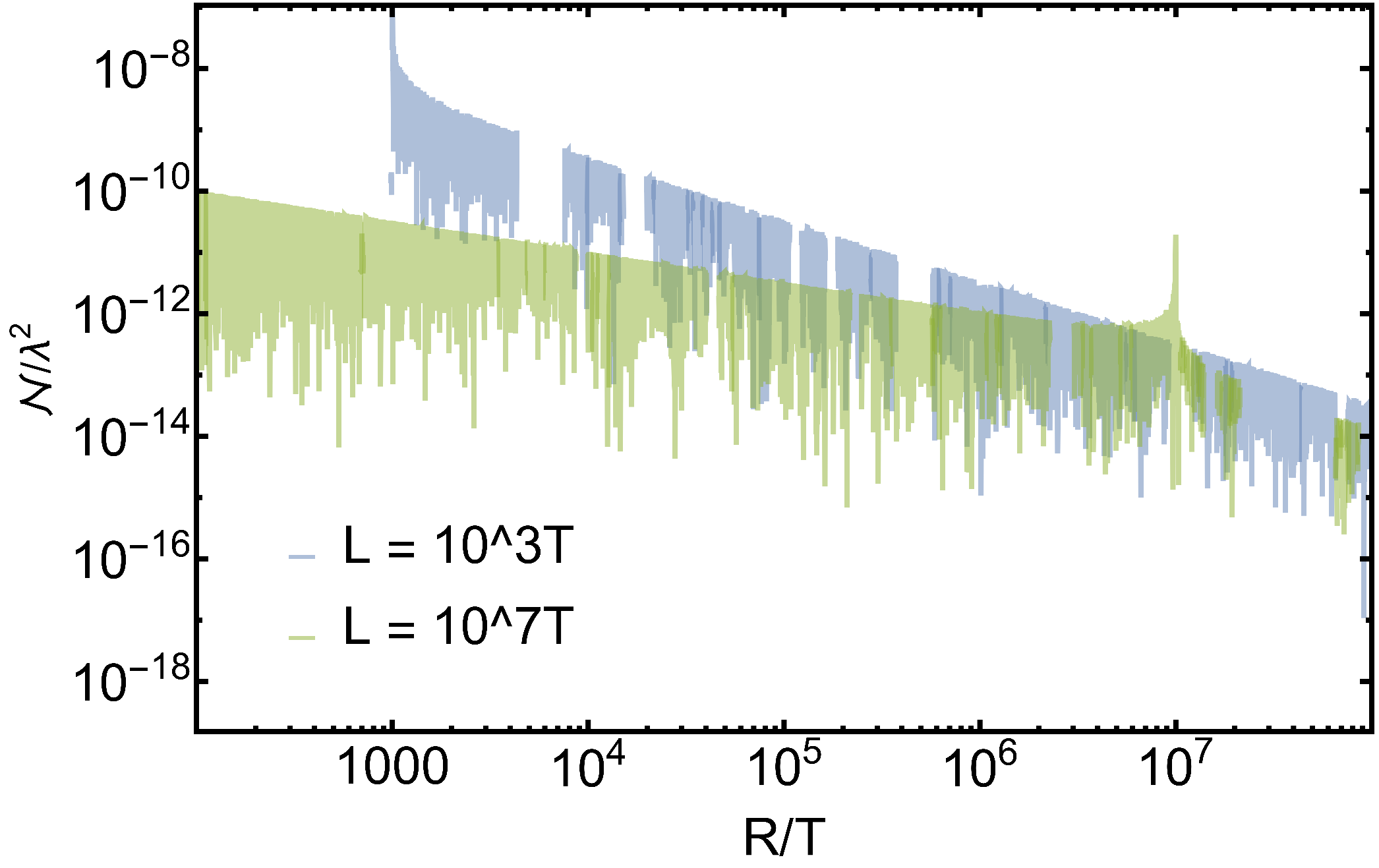} \caption{Negativity as a function of
gravitational waves source distance $R$ for multiple values of the detectors
separation $L$. We fixed the energy gap of the detectors as $\Omega T = 0.77$
for each of the plots. Other parameters is taken as $\sigma/T=1000$.}%
\label{Fig4}%
\end{figure}

In Fig. 4, we present how the entanglement harvested by detectors varies with
the distance from the source of gravitational waves, taking into account
different distances between the detectors themselves. We find that the
harvested entanglement diminishes as the source distance increases, a
phenomenon linked to the weakening strength of gravitational waves as they
propagate from their origin.

Additionally, in Fig. 5, we explore the relationship between the entanglement
harvested by the detectors and their energy gap across various source
distances. The analysis confirms that as the distance from the source
increases, the entanglement decreases, underscoring the impact of
gravitational wave attenuation over distance.

\begin{figure}[h]
\includegraphics[width=8.6cm]{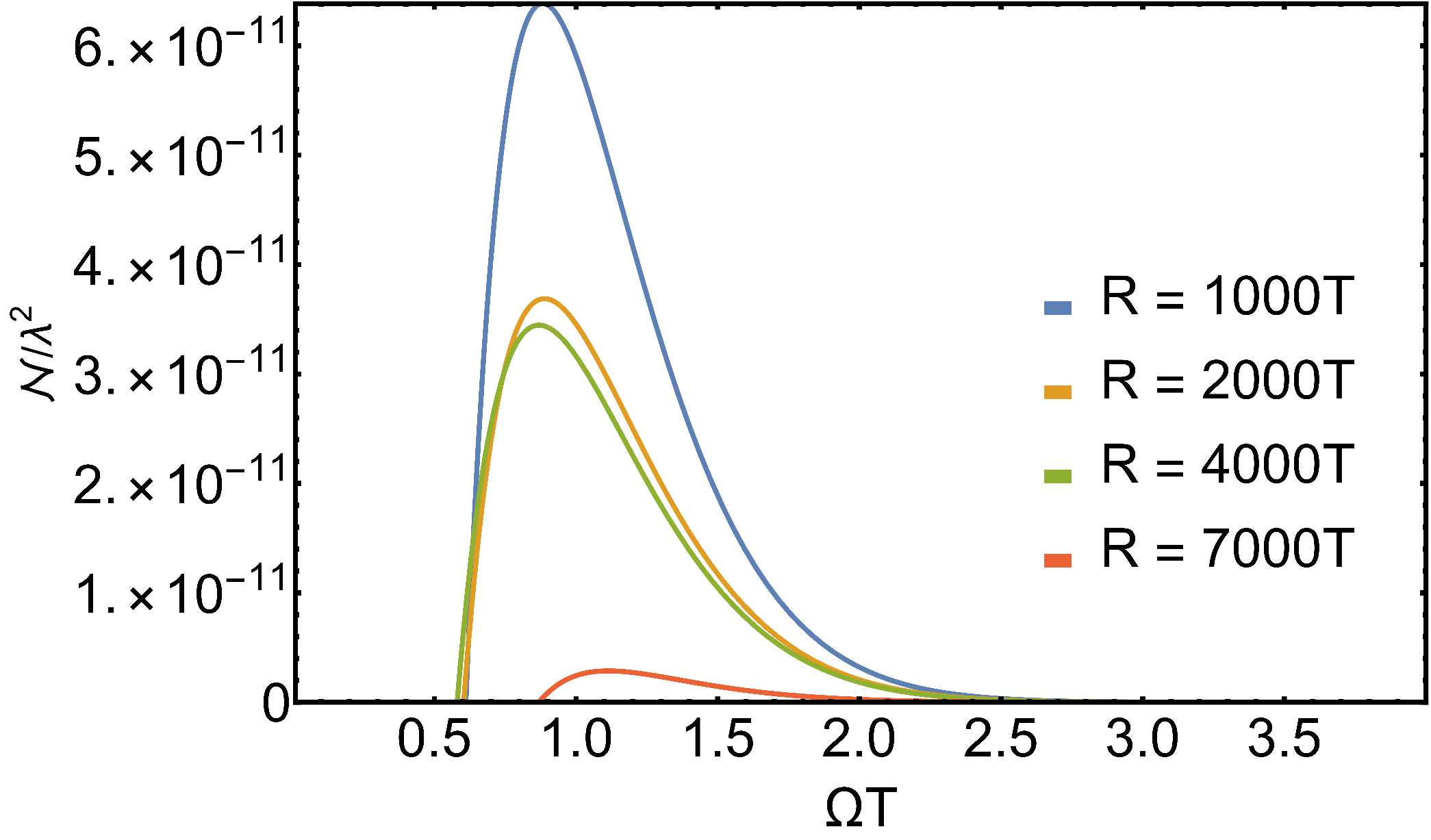} \caption{{Negativity as a function of
energy gap for multiple values of the source distance $R$. We fixed the
detector separation $L=10T$ as for each of the plots. Other parameters is
taken as $\sigma/T=1000$. }}%
\label{Fig5}%
\end{figure}

Finally we comment on realistic scales for the entanglement that can be
harvested by a physical system interacting with the cylindrical gravitational
field. Our plots for the negativity yielded (at best) $\mathcal{N}^{G}%
\sim\lambda^{2}5\ast10^{-7}$. Recall that the dimensionless coupling constant
$\lambda$ is given by $\sqrt{\frac{\pi}{2}}m/m_{p}$. If the mass of the system
is of the order of the mass of a hydrogen atom we would have $\lambda^{2}%
\sim10^{-38}$, so that the harvested negativity gives $\mathcal{N}^{G}%
\sim10^{-46}$. This result demonstrates that the entanglement harvested using
cylindrical symmetry is significantly greater than that observed in
entanglement harvesting from the gravitational field using hydrogen-like atoms
at a source distance of $R=1000T$ , as reported in \cite{prm23}. Furthermore,
even at a much larger distance of $R=9\times10^{7}T$ , the harvested
entanglement, $\mathcal{N}^{G}\approx10^{-55}$ , still exceeds previous
findings by 16 orders of magnitude. These findings show the enhanced
capability of detectors with cylindrical symmetry to harvest entanglement from
the GW field.

\subsection{Quantum Fisher Information}

In order to understand the information about the distance from the sources,
carried by the cylindrical GWs, we apply the concept of quantum Fisher
information (QFI) to investigate the possibility of measurement.

According to the quantum Cram\'{e}r-Rao theorem \cite{bc94,bcm96}, for a given
observable source distance $R$, the measurement precision is determined by
\begin{equation}
Var(R)\geq\frac{1}{n\mathcal{F}_{Q}(R)},
\end{equation}
where $Var$ is the covariant variance, and $n$ represents the number of
repeated measurements. $\mathcal{F}_{Q}$ is the QFI defined by~\cite{mgp09}
\begin{equation}
\mathcal{F}_{Q}(R)=\frac{\left(  \partial_{R}(1-\dot{\mathcal{L}^{G}})\right)
^{2}}{\dot{\mathcal{L}^{G}}}+\frac{\left(  \partial_{R}\dot{\mathcal{L}^{G}%
}\right)  ^{2}}{\dot{\mathcal{L}^{G}}}=2\frac{\left(  \partial_{R}%
\dot{\mathcal{L}^{G}}\right)  ^{2}}{\dot{\mathcal{L}^{G}}},
\end{equation}
then we have the transition rate as
\begin{align}
\dot{\mathcal{L}^{G}}  &  =\lambda^{2}\frac{8T\sigma^{8}}{15\pi}\int
d|\boldsymbol{k}||\boldsymbol{k}|^{10}J_{0}(\boldsymbol{k}R)J_{0}%
(\boldsymbol{k}(R-L))\nonumber\\
&  \quad\times e^{-|\boldsymbol{k}|^{2}\sigma^{2}}e^{-T^{2}(|\boldsymbol{k}%
|+\Omega)^{2}}\nonumber\\
&  \quad-\lambda^{2}\frac{8T^{3}\sigma^{8}}{15\pi}\int d|\boldsymbol{k}%
||\boldsymbol{k}|^{10}\nonumber\\
&  \quad\times J_{0}(\boldsymbol{k}R)J_{0}(\boldsymbol{k}%
(R-L))e^{-|\boldsymbol{k}|^{2}\sigma^{2}}\nonumber\\
&  \quad\times(|\boldsymbol{k}|+\Omega)^{2}e^{-T^{2}(|\boldsymbol{k}%
|+\Omega)^{2}}\nonumber\\
&  =\lambda^{2}\frac{8T\sigma^{8}}{15\pi}\int d|\boldsymbol{k}||\boldsymbol{k}%
|^{10}\nonumber\\
&  \quad\times J_{0}(\boldsymbol{k}R)J_{0}(\boldsymbol{k}%
(R-L))e^{-|\boldsymbol{k}|^{2}\sigma^{2}}\nonumber\\
&  \quad\times e^{-T^{2}(|\boldsymbol{k}|+\Omega)^{2}}\left[  1-T^{2}%
(|\boldsymbol{k}|+\Omega)^{2}\right]  .
\end{align}
and
\begin{align}
\partial_{R}\dot{\mathcal{L}^{G}}  &  =\lambda^{2}\frac{8T\sigma^{8}}{15\pi
}\int d|\boldsymbol{k}||\boldsymbol{k}|^{11}\left[  -J_{1}(\boldsymbol{k}%
R)J_{0}(\boldsymbol{k}(R-L))\right. \nonumber\\
&  \quad\left.  -J_{0}(\boldsymbol{k}R)J_{1}(\boldsymbol{k}(R-L))\right]
e^{-|\boldsymbol{k}|^{2}\sigma^{2}}e^{-T^{2}(|\boldsymbol{k}|+\Omega)^{2}%
}\nonumber\\
&  \quad\times\left[  1-T^{2}(|\boldsymbol{k}|+\Omega)^{2}\right]  .
\end{align}
Finally, we obtained the expression for the QFI as
\begin{align}
\mathcal{F}_{Q}(R)  &  =2\lambda^{2}\frac{8T\sigma^{8}}{15\pi}\int
d|\boldsymbol{k}||\boldsymbol{k}|^{11}\left(  -J_{1}(\boldsymbol{k}%
R)J_{0}(\boldsymbol{k}(R-L))\right. \nonumber\\
&  \left.  -J_{0}(\boldsymbol{k}R)J_{1}(\boldsymbol{k}(R-L))\right)
\nonumber\\
&  \times e^{-|\boldsymbol{k}|^{2}\sigma^{2}}e^{-T^{2}(||\boldsymbol{k}%
|+\Omega)^{2}}\left(  1-T^{2}(|\boldsymbol{k}|+\Omega)^{2}\right)
\end{align}

\begin{figure}[h]
\includegraphics[width=8.6cm]{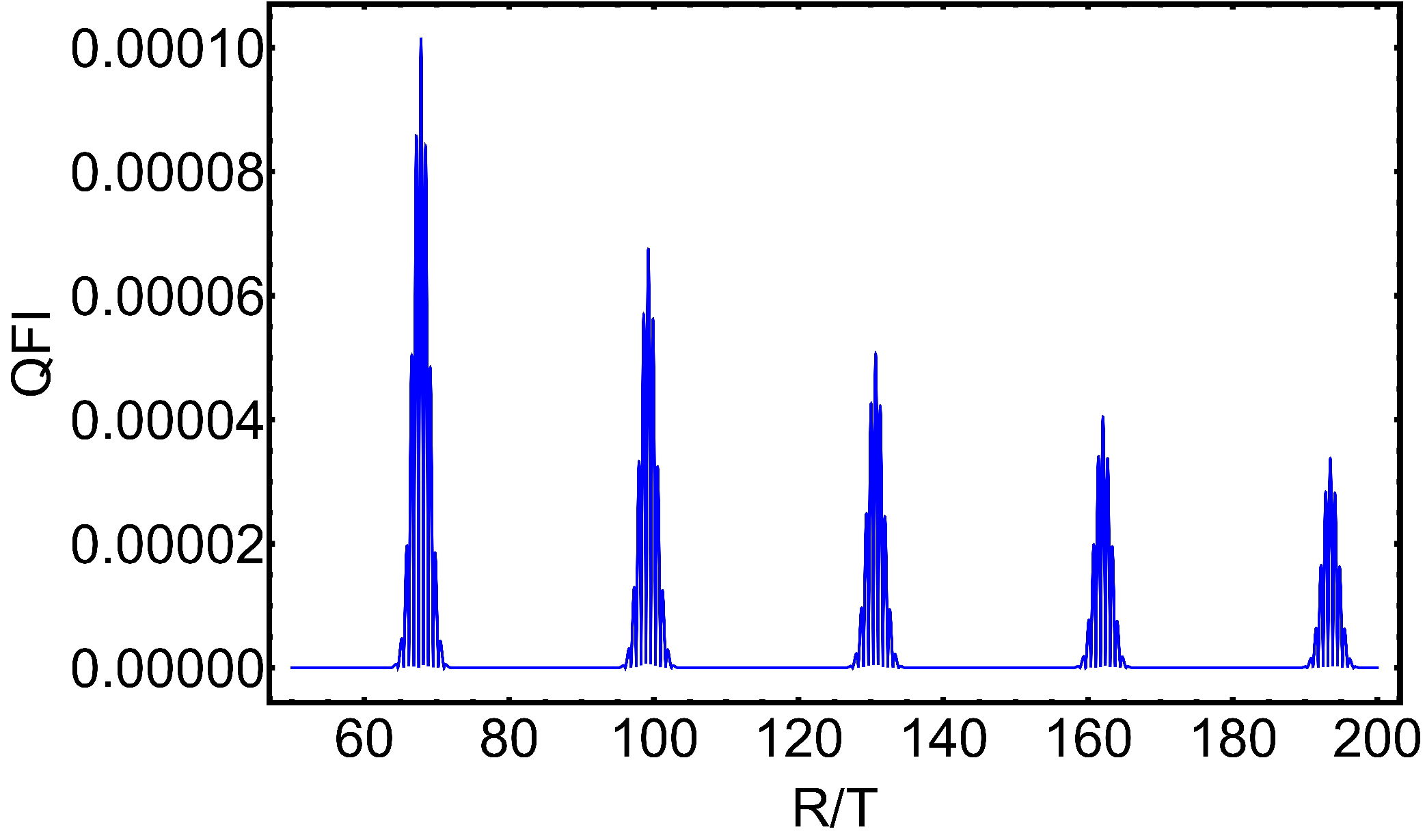} \caption{{The plot displays the
absolute value of the QFI as a function of the normalized wave source distance
. The parameters used are $\sigma= 0.3T$, $\Omega T = 0.1$, and $L = 10T$.}}%
\label{Fig6}%
\end{figure}

The measurement uncertainty is defined by
\begin{equation}
U_{R}=\frac{\sigma_{R}}{R},
\end{equation}
where $\sigma_{R}=\sqrt{Var(R)}$. We found that, for our model, when the wave
source distance satisfies $R/L<200$, the uncertainty in measuring the wave
source distance is approximately $21\%$. Moreover, the QFI decreases rapidly
as $R$ increases as shown in Fig. 6. This level of uncertainty is comparable
to LIGO's measurements of binary neutron star merger events ($10\%-20\%$) and
is lower than the uncertainty in measurements of binary black hole merger
events ($20\%-50\%$)~\cite{aaa17,nhs10}. Moreover, it is noted that Fig. 6
exhibits abrupt, non-smooth behavior, which is derived from the oscillatory
Bessel functions. These functions, combined with the exponential suppression
term, introduce a resonance effect that leads to abrupt changes in the QFI.

\section{Conclusion}

In this paper, we have explored the entanglement harvesting protocol within
the spacetime in the presence of cylindrical gravitational waves, revealing
results that markedly contrast with those from scenarios involving standard
quantum gravitational field. The magnitude of entanglement negativity is
substantially greater than that harvested from the vacuum of a conventional
gravitational field. Importantly, our research elucidates the relationship
between entanglement harvesting and the source distance of gravitational waves.

This significant discrepancy highlights the unique quantum structure and
entanglement properties of the vacuum state associated with cylindrical GWs.
It indicates that these specialized gravitational configurations may be
particularly effective for entanglement harvesting, potentially enabling the
observation of the information about the distance from sources at scales much
larger than previously thought possible.

\acknowledgments

This work is supported by National Natural Science Foundation of China (NSFC)
with Grant No. 12375057 and the Fundamental Research Funds for the Central
Universities, China University of Geosciences (Wuhan).

\end{document}